\documentclass[11pt]{article}
\usepackage{amssymb, amsmath}
\usepackage{amsthm}
\usepackage{mathabx}	

\newtheorem{proposition}{Proposition}
\newtheorem{theorem}{Theorem}
\newtheorem{corollary}{Corollary}

\usepackage{mathptmx}       
\usepackage{helvet}         
\usepackage{courier}        
\usepackage{type1cm}        
\usepackage{verbatim}
\usepackage{makeidx}         
\usepackage{graphicx}        
\usepackage{multicol}        
\usepackage[bottom]{footmisc}

\usepackage{url}

\author{
  Klaus Wehmuth\\
  National Laboratory for Scientific Computing (LNCC)\\
  Av. Get\'{u}lio Vargas, 333\\
  25651-075 -- Petr\'{o}polis, RJ -- Brazil\\
   \texttt{klaus@lncc.br}
  \and
  \'{E}ric Fleury\\
    LIP -- UMR CNRS 5668\\
    ENS de Lyon, INRIA, CNRS, Universit\'{e} Lyon 1\\
    69364 Lyon, France\\
  \texttt{eric.fleury@ens-lyon.fr}
  \and
  Artur Ziviani\\
  National Laboratory for Scientific Computing (LNCC)\\
  Av. Get\'{u}lio Vargas, 333\\
  25651-075 -- Petr\'{o}polis, RJ -- Brazil\\
  \texttt{ziviani@lncc.br}
}
\title{On MultiAspect Graphs}
\date{}

\begin{document}
\maketitle

\begin{abstract}
Different graph generalizations have been recently used in an ad-hoc manner to represent multilayer networks, \emph{i.e.} systems formed by distinct layers where each layer can be seen as a network. Similar constructions have also been used to represent time-varying networks.
We introduce the concept of MultiAspect Graph (MAG) as a graph generalization that we prove to be isomorphic to a directed graph, and also capable of representing all  previous generalizations.
In our proposal, the set of vertices, layers, time instants, or any other independent features are considered as an \emph{aspect} of the MAG. 
For instance, a MAG is able to represent multilayer or time-varying networks, while both concepts can also be combined to represent a multilayer time-varying network and even other higher-order networks.
Since the MAG structure admits an arbitrary (finite) number of aspects, it hence introduces a powerful modelling abstraction for networked complex systems.
This paper formalizes the concept of MAG and derives theoretical results useful in the analysis of complex networked systems modelled using the proposed MAG abstraction.
We also present an overview of the MAG applicability.
\end{abstract}

%
%



\section{Introduction}
\label{sec:Int}
Many graph generalizations have been proposed in the related literature, e.g. hypergraphs~\cite{Distel2010}, with the purpose of representing edges with multiple vertices, so that every vertex on an edge is related to all other vertices in that edge. In particular, k-uniform hypergraphs are sometimes referred to as k-graphs or generalized graphs~\cite{Erdos1964, Bollobas1965, Chvatal1971}. 

More recently, other graph generalizations have been proposed for modelling systems that are described as the conjunction of distinct interdependent networks, where each of these networks can be seen as distinct layers, each of which can be represented by a graph. These generalizations, usually known as multilayer networks~\cite{Kurant2006a, DeDomenico2013, Gomez2013, Kivela14072014}, propose structures where (some) vertices may be connected in distinct layers. 
As an example, consider an urban multimodal public transportation system. This system can be modelled by distinct layers, such as the bus network, the tramway network, the subway network, the commuter rail network, and so on. Note that, in such an arrangement, each layer can be modelled by an independent graph and some vertices~(\emph{i.e.}, bus and tramway stops; commuter rail and metro stations) may be shared by distinct layers, thus connecting the whole system. Similar graph generalizations have been proposed for time-varying networks~\cite{Casteigts2012,Holme2012,flocchini2013,holme2013,Wehmuth2015a}, where the structure of the network (\emph{e.g.} vertices or edges) may vary in time. Although these models are in use for some time, they are not general enough to be able to combine these two features and represent a time-varying multilayer network.
An example of such a network is a multimodal transportation system where, as before, the layers represent the distinct transportation modes and the vertices are the stops, but the time may be added to represent the service schedule. In this network, an edge is constructed by six parameters, two representing vertices, two representing layers, and two representing time instants. For instance, a bus leaving the bus stop $b_1$ at time $t_a$ and arriving to the stop $b_2$ at time $t_b$ can be represented by the edge $(b_1, bus, t_a, b_2, bus, t_b)$. 
There is a lack of a formal unified representation for such complex networked systems.

In this paper, we formalize the concept of MultiAspect Graph (MAG)\footnote{Disambiguation: Note that MAG is also an acronym used in~\cite{Kim2012-mag} for ``Multiplicative Attribute Graph''. The Multiplicative Attribute Graph model intends to model networks in which nodes may have attributes and the proposed model captures the interactions between the network structure and the node attributes, which is a completely different purpose than the MultiAspect Graph (MAG) abstraction introduced in this paper.} as a further generalization able to represent multilayer and time-varying networks, as well as time-varying multilayer networks and even other higher-order networks. 
A similar idea was briefly discussed in~\cite{Mucha2010b}, whereas without a thorough formalization. In a MAG, the set of vertices, layers, time instants, or any other independent feature, is considered as an \emph{aspect}. On a MAG with $p$ aspects, an edge is a $2p$-tuple, formed by two elements of each aspect, grouped in two parts, each containing one element of each aspect. For instance, a multilayer network can be represented by a MAG with $2$ aspects, the first being the set of vertices and the second the set of layers. Each edge in this example is a quadruple, containing $2$ vertices and $2$ layers. Note that a time-varying multilayer network,
such as the previously discussed multimodal transportation system, is namely a MAG with $3$ aspects.

We formally define a MAG and its basic properties. The key contribution of this paper is to show that a MAG is closely related to a traditional oriented graph. This relation can be used to analyze properties and applications of the MAG concept on network analysis. We further discuss some general examples of the applicability of order 2 and 3 MAGs as well as of the basic MAG representation and algorithms.
It is our intent to make this work accessible to a broader audience, which might be interested in modelling 
and analyzing complex networked systems, where the MAG formalization introduced in this paper can prove helpful. Consequently,  we adopt a relatively verbose proof and presentation style in an effort to increase the readability by a broader audience.

The remainder of this work is structured as follows. 
Section~\ref{sec:mag} introduces the concept of a MultiAspect Graph~(MAG).
 The main properties of a MAG are derived in Section~\ref{sec:magprop}.
Section~\ref{sec:app} discusses the applicability of the MAG abstraction.
Section~\ref{sec:conc} concludes the paper and discusses possible future works.

\section{MultiAspect Graph (MAG)}
\label{sec:mag}

In this section, we introduce the concept of a MultiAspect Graph~(MAG).

\subsection{MAG definition}
\label{subsec:magdef}

We define a MAG as $H = (A, E)$, where $E$ is a set of edges and $A$ is a finite list of sets, each of which is called an \emph{aspect}. Each aspect $\sigma \in A$ is a finite set, and the number of aspects $p = |A|$ is called the order of $H$. Each edge $e \in E$ is a tuple with $2 \times p $ elements. All edges are constructed so that they are of the form $(a_1,\dots,a_p,  b_1,\dots,b_p)$, where $a_1, b_1$ are elements of the first aspect of $H$, $a_2,b_2$ are elements of the second aspect of $H$, and so on, until $a_p, b_p$ which are elements of the $p$-th aspect of $H$.

As a matter of notation, we say that $A(H)$ is the aspect list of $H$ and $E(H)$ is the edge set of~$H$. Further, $A(H)[n]$ is the $n$-th aspect in $A(H)$,  $|A(H)[n]| = \tau_n$ is the number of elements in $A(H)[n]$, and $|A(H)|$ is the order of $H$. 
We adopt the convention of calling the elements of the first aspect of a MAG ($A(H)[1]$) as vertices, and also, the convention of calling the first $p$ entries of the edge $e$ as the origin elements of $e$, and the last $p$ entries as the destination elements of the edge $e$. 
Further, we define the following two sets constructed from the cartesian products of aspects of an order $p$ MAG:
\begin{equation}
\mathbb{V}(H) = \bigtimes_{n=1}^p A(H)[n],
\end{equation}
the cartesian product of all the aspects of the MAG $H$, and
\begin{equation}
\mathbb{E}(H) =  \bigtimes_{n=1}^{2p} A(H)[(n-1)(mod \ p)+1],
\end{equation}
which is the set of all possible edges in the MAG $H$, so that $E(H) \subseteq \mathbb{E}(H)$. 

We call $\mathbf{u} \in \mathbb{V}(H)$ a \emph{composite vertex} of MAG $H$. As a matter of notation, a composite vertex is always represented as a bold lowercase letter, as in $\mathbf{u}$, for instance.
By construction, a pair $(\mathbf{u},\mathbf{v})$ of composite vertices is closely related to an edge $e \in \mathbb{E}(H)$. In fact, it can be seen that for every edge $e \in \mathbb{E}(H)$ there are two composite vertices $\mathbf{u},\mathbf{v} \in \mathbb{V}(H)$ such that the entries of $\mathbf{u}$ match the origin elements of $e$ and the entries of $\mathbf{v}$ match the destination elements of $e$. 
From this, for an edge $e = (a_1,a_2,\dots,a_p, b_1,b_2,\dots,b_p) \in E(H)$ we can define the functions
\begin{align}
\pi_o:\mathbb{E}(H)   & \to \mathbb{V}(H) \\
e = (a_1,a_2,\dots,a_p, b_1,b_2,\dots,b_p) & \mapsto  (a_1,a_2,\dots,a_p), \notag
\end{align}
which maps the origin elements of an edge to a composite vertex, and
\begin{align}
\pi_d:\mathbb{E}(H) & \to  \mathbb{V}(H) \\
e = (a_1,a_2,\dots,a_p, b_1,b_2,\dots,b_p) & \mapsto (b_1,b_2,\dots,b_p), \notag
\end{align}
which maps the destination elements of an edge to a composite vertex.
Based on these two functions, we also define the bijective function
\begin{align}
\label{func:psi}
\psi: \mathbb{E}(H)& \to \mathbb{V}(H) \bigtimes \mathbb{V}(H) \\
e  & \mapsto  (\pi_o(e), \pi_d(e)) = ((a_1,a_2,\dots,a_p),  (b_1,b_2,\dots,b_p)). \notag
\end{align}

Similarly to a simple graph, we do not allow the presence of self-loop edges on a MAG, \emph{i.e.} for any MAG $H$, if $e_\ell \in \mathbb{E}(H)$ is a self-loop edge, then $e_\ell \notin E(H)$. In a MAG, a self-loop edge is an edge $e_\ell \in \mathbb{E}(H)$, such that $\pi_o(e_\ell) = \pi_d(e_\ell)$. 

Further, since on an order $p$ MAG an edge is a tuple with $2 \times p$ elements, we also define $2 \times p $ canonical projections, where each of those projections maps an edge to one of its elements:
\begin{align}
\pi_{n}: E(H) & \to A(H)[n]  \\
(a_1,\dots,a_p,  b_1,\dots,b_p) & \mapsto a_n, \notag \\
& \notag \\
\pi_{p+n}: E(H) & \to A(H)[n] \\
(a_1,\dots,a_p,  b_1,\dots,b_p) & \mapsto b_n. \notag
\end{align} 

We also use the projections $\pi_1$ to $\pi_p$ defined above to recover each of the values that compose the $p$-tuple which characterizes a composite vertex. This is an abuse of notation, since the domain of these projections is a set of edges instead of a set of composite vertices, but the intuition and readability gained with this notation justifies its use.

\subsection{Aspect sub-determination}
\label{subsec:subdt}

The sub-determination of the composite vertices in a MAG $H$ partitions $\mathbb{V}(H)$ into equivalence classes considering only a partial aspect sublist. As a consequence, the sub-determination of the composite vertices leads to edge and MAG sub-determinations. For instance, the traditional aggregated direct graph, commonly found in the time-varying graph literature, is a particular case of MAG sub-determination. We detail this in the following subsections.

\subsubsection{Sub-determined composite vertices}
\label{subsubsec:subdetvt}
On a given MAG $H$, we use a nonempty proper sublist\footnote{A proper sublist of a given list~$L$ is a sublist which is strictly contained in $L$ and so necessarily excludes at least one member of $L$.} of the aspects of the MAG to characterize an equivalence class, which is then used to partition the set of composite vertices.

For a MAG $H$ of order $p$, there are $2^p - 1$ proper sublists of $A(H)$. As we require the sublist of aspects used to characterize an equivalence class to be nonempty, it follows that it can be characterized in $2^p - 2$ distinct ways. For each of these $2^p - 2$ ways, we have a list $A_C(H) \subset A(H)$ of the aspects used to determine an equivalence class.
Note that in a MAG of order $p = 1$ (\emph{i.e.} a traditional graph), a vertex can not be sub-determined, since $2^p - 2 = 0$.

Let $\zeta$, with $1 \leq \zeta \leq 2^p - 2$, be an index for one of the possible ways to construct a proper nonempty sublist of aspects. 
From this, we can define a canonical representation of the sub-determination directly defined by $\zeta$. For any given $\zeta$, we consider the $p$-bit binary expansion of $\zeta$ that is used as an indicator showing which aspects of the original MAG are present on the sub-determination.
More specifically, the least significant bit indicates the presence or absence of the first aspect and the most significant bit indicates the presence or absence of the last aspect. By this convention, in a MAG with $p=3$ aspects, we have that $\zeta = \texttt{001}_2$ corresponds to the sub-determination where only the first aspect is present, $\zeta = \texttt{010}_2$ corresponds to the sub-determination where only the second aspect is present, $\zeta = \texttt{101}_2$ corresponds to the sub-determination where both the first and the third aspects are present, and so on.
By using this convention, we can directly associate a given $\zeta$ to its corresponding aspect sublist.

Therefore, for each $\zeta$, we have a unique sublist $A_{\zeta}(H)$ of aspects, such that $p_\zeta = |A_{\zeta}(H)|$ is the order of the sub-determination $\zeta$. 
We now define the set 
\begin{equation}
\mathbb{V}_\zeta(H) = \bigtimes_{n=1}^{p_\zeta} A_{\zeta}(H)[n], 
\end{equation}
where $\mathbb{V}_\zeta(H)$ is the cartesian product of all the aspects in the sublist $A_{\zeta}(H)$ of aspects, according to the index $\zeta$.
We call $\mathbf{u}_\zeta \in \mathbb{V}_\zeta(H)$ a sub-determined vertex, according to the sub-determination $\zeta$.

We can now define the function
\begin{align}
S_\zeta: \mathbb{V}(H) & \to \mathbb{V}_\zeta(H)  \\
(a_1,a_2,\dots,a_p) & \mapsto (a_{\zeta_1}, a_{\zeta_2},\dots,a_{\zeta_m}), \notag
\end{align}
where $m=p_\zeta$.  $S_\zeta$ maps a composite vertex $\mathbf{u} \in \mathbb{V}(H)$ to the corresponding sub-determined composite vertex $\mathbf{u}_\zeta \in \mathbb{V}_\zeta(H)$, according to the sub-determination $\zeta$. As $(a_{\zeta_1}, a_{\zeta_2},$ $\dots,a_{\zeta_m})$ $\in \mathbb{V}_\zeta(H)$, it follows that $a_{\zeta_1} \in A_{\zeta}(H)[1], \dots, a_{\zeta_m}\in A_{\zeta}(H)[m]$.
From the definition, it can be seen that the function $S_\zeta$ is not injective. Hence, the function $S_\zeta$ for a given sub-determination can be used to define a equivalence relation $\equiv_\zeta$ in $\mathbb{V}(H)$, where for any given composite vertices $\mathbf{u}, \mathbf{v} \in \mathbb{V}(H)$, we have that $\mathbf{u} \equiv_\zeta \mathbf{v}$ if and only if $S_\zeta(\mathbf{u}) = S_\zeta(\mathbf{v})$.

\subsubsection{Sub-determined edges}
\label{subsubsec:subdetedg}
From the  sub-determination $\zeta$ of order $p_\zeta$, we can also construct the set
\begin{equation}
\mathbb{E}_\zeta(H) =  \bigtimes_{n=1}^{2 \times p_\zeta} A_{\zeta}(H)[(n-1)(mod \ p_\zeta)+1], 
\end{equation}
where $ p_\zeta =|A_{\zeta}(H)|$ is the order of the sub-determination $\zeta$, and $\mathbb{E}_\zeta(H)$ is the set of all possible sub-determined edges according to $\zeta$. We then define the function
\begin{align}
E_\zeta: \mathbb{E}(H) & \to \mathbb{E}_\zeta(H) \\
(a_1,a_2,\dots,a_p, b_1,b_2,\dots,b_p) & \mapsto (a_{\zeta_1}, a_{\zeta_2},\dots,a_{\zeta_m}, b_{\zeta_1}, b_{\zeta_2},\dots,b_{\zeta_m}), \notag
\end{align}
where $m=p_\zeta$ and $a_{\zeta_1},b_{\zeta_1} \in A_{\zeta}(H)[1], a_{\zeta_2},2_{\zeta_2} \in A_{\zeta}(H)[2], \dots, a_{\zeta_m},b_{\zeta_m} \in A_{\zeta}(H)[m]$. This function takes an edge to its sub-determined form according to $\zeta$ in a similar way as defined above for composite vertices.
In general, the function $E_\zeta$ is not injective. Consider two distinct edges $e_1,e_2 \in E(H)$, such that $e_1$ and $e_2$ differ only in aspects which are not in $A_{\zeta}(H)$. Since $E_\zeta(\cdot)$ only contains values for aspects present in $A_{\zeta}(H)$, it follows that $E_\zeta(e_1) = E_\zeta(e_2)$, and therefore $E_\zeta$ is not injective. Further, consider an edge $e \in E(H)$ and its sub-determined edge $e_\zeta = E_\zeta(e)$, such that $\pi_o(e_\zeta) = \pi_d(e_\zeta)$, \emph{i.e.} $e_\zeta$ is a self-loop. Since self-loops are not allowed to be present on a MAG, it follows that $e_\zeta \notin E_\zeta(E(H))$. 
As consequence, we have that $|E_\zeta(E(H))| \leq |E(H)|$.

\subsubsection{Sub-determined MAGs}
\label{subsubsec:subdetmag}
For a given sub-determination $\zeta$ we have the sublist $A_{\zeta}(H)$ of considered aspects and also the sub-determined edges obtained from $\zeta$. Based on them, we can now obtain a sub-determined MAG. For a given sub-determination $\zeta$ we define the function
\begin{align}
M_\zeta: (A(H), E(H)) & \to (A_{\zeta}(H), \mathbb{E}_\zeta(H)) \\
H & \mapsto (A_{\zeta}(H), E_\zeta(E(H))). \notag
\end{align}
Since $A_{\zeta}(H)$ is the sublist of aspects of $H$ prescribed by $\zeta$ and $E_\zeta(E(H))$ is the set of all sub-determined edges according to the sub-determination $\zeta$, it follows that $(A_{\zeta}(H), E_\zeta(E(H)))$ is a MAG obtained from $H$ according to the sub-determination $\zeta$.
As $|A_{\zeta}(H)| < |A(H)|$, it follows that the order of $M_\zeta(H)$ is lower than the order of $H$. Further, since self-loops may be created by edge sub-determination and discarded, and also since $E_\zeta$ is not injective, it follows that $|E_\zeta(E(H))| \leq |E(H)|$.

\subsubsection{Aggregated directed graph}
\label{subsubsec:aggr}
The concept of aggregated graph is usually found in the literature associated with time-varying or multilayer networks. In these environments, it consists of ignoring the time and layer aspects while projecting all edges over the vertices set. Since MAGs can be used to represent time-varying and multilayer graphs, we present a similar concept extended to the MAG environment.
 
The aggregated  graph associated with a MAG is a directed  graph created by a particular case of MAG sub-determination, $\zeta = 1$, where $A_1(H) = [A(H)[1]]$, the sublist of $H$ which contains only the first aspect of the MAG $H$. In this case, for a given MAG $H$ with $p$ aspects, we have
\begin{align}
E_{1}: \mathbb{E}(H) & \to A(H)[1] \bigtimes A(H)[1]  \\
(a_1,a_2,\dots,a_p, b_1,b_2,\dots,b_p) & \mapsto (a_1,b_1), \notag
\end{align}
where $a_1,b_1 \in A(H)[1]$, and
\begin{align}
M_{1}: (A(H), E(H)) & \to ([A(H)[1]], A(H)[1] \bigtimes A(H)[1])  \\
H & \mapsto ([A(H)[1]], E_{1}(E(H))). \notag
\end{align}
Note that since $M_{1}(H)$ is an order 1 MAG, it follows that $M_{1}(H)$ is a traditional directed graph, as every $a \in A(H)[1]$ is a vertex, and every edge $e \in E_{1}(E(H)) \subseteq A(H)[1] \bigtimes A(H)[1]$.

\section{MAG properties}
\label{sec:magprop}

In this section, we derive the main properties of a MAG as follows.
Section~\ref{subsec:magiso} discusses isomorphisms between MAGs. Section~\ref{subsec:isogrph} presents isomorphisms between MAGs and traditional directed graphs. The theoretical results obtained in Section~\ref{subsec:isogrph} are key for the subsequent subsections, where the relation between MAGs and directed graphs is further explored. Section~\ref{subsec:degree}  defines the concept of degree on a MAG. Section~\ref{subsec:adj} explores adjacency on MAGs and its relation to adjacency in directed graphs. Section~\ref{subsec:wtpc} discusses the relations between walks, trails, paths, and cycles on a MAG and on a traditional directed graph. Sections~\ref{subsec:shortph} discusses shortest paths on MAGs.

\subsection{MAG isomorphism}
\label{subsec:magiso}
Two MAGs of order $p$, H and K, are isomorphic if $p = |A(H)| = |A(K)|$, and there are $p$ bijective functions $f_n: A(H)[n] \to A(K)[n]$ such that $(a_1,a_2,\dots,a_p,  b_1,$ $b_2,\dots,b_p) \in E(H)$ if and only if $(f_1(a_1),\dots,f_p(a_p), f_1(b_1),\dots,f_p(b_p))$ $\in E(K)$, where $a_1,b_1 \in A(H)[1]$, $a_2,b_2 \in A(H)[2], \dots , a_p,b_p \in A(H)[p]$. 

Since the MAG isomorphism is an equivalence relation, the set of all MAGs isomorphic to a given MAG $H$ form an equivalence class in the set of all MAGs. This equivalence relation partitions the set of all MAGs.
Further, since the functions $f_p$ are bijections, it follows that if two MAGs $H$ and $K$ are isomorphic, they necessarily are of the same order, and each pair of aspects in $H$ and $K$ has the same number of elements, \emph{i.e.} $|A(H)| = |A(K)|$ and $|A(H)[n]| = |A(K)[n]|$, for all $0 < n \leq p$. 
From the requirement that an  edge $(a_1,a_2,\dots,a_p,  b_1,b_2,\dots,b_p)$ exists in $H$ if and only if the  edge $(f_1(a_1),f_2(a_2),\dots,f_p(a_p), f_1(b_1),f_2(b_2),\dots,f_p(b_p))$ exists in $K$, it can be seen that two isomorphic MAGs also have the same number of  edges, \emph{i.e.} $|E(H)| = |E(K)|$.

In addition to the isomorphism between MAGs, a  special case of the MAG isomorphism can be constructed, which characterizes a natural isomorphism (\emph{i.e.} a natural choice of isomorphism) between a MAG $H$ and a directed graph $G$. To achieve this, we make the directed  graph $G$ such that its vertex set $V(G)$ is equal to the set $\mathbb{V}(H)$ and use the identity function $I:\mathbb{V}(H) \to V(G)$ as the bijective function to characterize the isomorphism. 

In Section~\ref{subsec:isogrph}, we present Theorem~\ref{theo:iso} with a more general proof of the existence and uniqueness of the isomorphism between a MAG and a directed graph. Note that the idea of the natural isomorphism we just mentioned can provide an intuitive insight for the proof of Theorem~\ref{theo:iso}.

\subsection{Isomorphism between MAGs and directed graphs}
\label{subsec:isogrph}
We say a MAG $H$ is isomorphic to a traditional directed graph G when there is a bijective function $f:\mathbb{V}(H) \to V(G)$, such that an edge $e \in E(H)$ if and only if the edge $(f(\pi_o(e)), f(\pi_d(e))) \in E(G)$. 

\begin{theorem}{For every MAG $H$ of order $p > 0$, where all aspects are non-empty sets, there is a unique (up to a graph isomorphism) directed graph G with $\prod_{n=1}^p \tau_n$ vertices which is isomorphic to the MAG $H$. Note that $\tau_n = |A(H)[n]|$ is the number of elements on the $n$-th aspect of $H$. }
\label{theo:iso}
\begin{proof}
We show that for any such MAG $H$ there is a unique (up to a graph isomorphism) directed graph with $\prod_{n=1}^p \tau_n$ vertices for which there is a bijective function $f:\mathbb{V}(H) \to V(G)$, such that any edge $e = (a_1,a_2,\dots,a_p,  b_1,b_2,\dots,b_p) \in E(H)$ if and only if $(f(\pi_o(e)), f(\pi_d(e))) = (f((a_1, a_2,\dots,a_p)), f((b_1, b_2,\dots,b_p))) \in E(G)$.
\begin{itemize}
\item Existence of $G$:\\
Given the MAG $H$, we construct a directed graph $G$ which satisfies the isomorphism conditions. We start with a directed graph $G$ with $\prod_{n=1}^p \tau_n$ vertices and no edges. Note that the number of vertices in $G$ equals the number of composite vertices in $H$, \emph{i.e.} $|V(G)| = |\mathbb{V}(H)| = \prod_{n=1}^p \tau_n$. We then take an arbitrary bijective function $f:\mathbb{V}(H) \to V(G)$. Since the sets $V(G)$ and $\mathbb{V}(H)$ have the same number of elements, such bijection exits. Finally, for every edge $e = (a_1,a_2,\dots,a_p,  b_1,b_2,\dots,b_p) \in E(H)$ we add an edge $(f(\pi_o(e)), f(\pi_d(e)))$ to $E(G)$. Since $f$ is injective, it follows that if $\pi_o(e)  \neq  \pi_d(e)$ then $f(\pi_o(e)) \neq f(\pi_d(e))$. Therefore, each distinct edge $e \in E(H)$ is mapped to a distinct edge $(f(\pi_o(e)), f(\pi_d(e)))\in E(G)$. As the only edges in $E(G)$ are the ones mapped from $E(H)$, it follows that the edge $e \in E(H)$ if and only if the edge $(f(\pi_o(e)), f(\pi_d(e))) \in E(G)$, as required. Note that, as a consequence, we have that $|E(H)| = |E(G)|$. This gives us a directed graph $G$ and a bijective function $f$ that satisfies the isomorphism requirements. Therefore, we have shown that the required directed graph $G$ exists.

\item Uniqueness of $G$:\\
Let's assume that in addition to the MAG $H$, the directed  graph $G$, and the bijective function $f$ described above, we also have another directed  graph $J$ with $\prod_{n=1}^p \tau_n$ vertices and a bijective function $j:\mathbb{V}(H) \to V(J)$, such that any edge $e \in E(H)$ if and only if the  edge $(j(\pi_o(e)), j(\pi_d(e))) \in E(J)$. 
Since both $f$ and $j$ are bijective functions, it follows that the composite function $(j  \circ f^{-1}):V(G) \to V(J)$ is also a bijection. Further, from the definitions of $f$ and $j$, it follows that the vertices $u, v \in V(G)$ are adjacent in $G$ if and only if the vertices $(j \circ f^{-1})(u), (j  \circ f^{-1})(v) \in V(J)$ are adjacent in $J$. The converse follows from the same argument applied to an edge in $E(J)$.
Therefore, $G$ and $J$ are isomorphic directed  graphs, and thus $G$ is unique up to a graph isomorphism.
\end{itemize}
Given the existence and uniqueness of the directed graph $G$, the existence of the function $f$, and since $H$ is an arbitrary MAG, we conclude that the theorem holds.
\end{proof}
\end{theorem}

As a matter of notation, hereafter we use $e$ to refer to MAG edges and $s$ to refer to traditional directed graph edges.

\begin{corollary}{Given a MAG $H$ and a directed graph $G$ isomorphic to $H$, there is a bijective function from $E(H)$ to $E(G)$, built upon the isomorphism characterised by the bijection $f$, and which takes each  edge $e \in E(H)$ to its corresponding edge $s \in E(G)$.}
\label{coro:bijct}
\begin{proof}
Let $H$ be a MAG and $G$ a directed graph isomorphic to $H$.
Since $H$ and $G$ are isomorphic, an edge $e$ belongs to $E(H)$ if and only if a corresponding edge $s= (f(\pi_o(e)), f(\pi_d(e)))$ belongs to $E(G)$, and the function $f$ is a bijection from $\mathbb{V}(H)$ to $V(G)$. Consider the following function
\begin{align}
\label{func:h}
h:E(H) & \to E(G)\\
e & \mapsto (f(\pi_o(e)), f(\pi_d(e))). \notag
\end{align}
Since $H$ and $G$ are isomorphic, it follows that if $e \in E(H)$ then $s = (f(\pi_o(e)), f(\pi_d(e)))$ $\in E(G)$. Further, since function $f$ is bijective, for every $s \in E(G)$ there is a unique $e \in E(H)$ such that $s = h(e)$. Therefore, we can conclude that $h$ is also bijective and the corollary holds.
\end{proof}
\end{corollary}

Theorem~\ref{theo:iso} as well as Corolary~\ref{coro:bijct} represent an important theoretical result because this allows the use of the isomorphic directed graph as a tool to analyze both the properties of a MAG and the behavior of dynamic processes over a MAG.

\subsubsection{MAG representation by composite vertices}
\label{subsubsec:crep}
We now show that it is possible to create a representation of any given MAG using composite vertices. This is equivalent to the natural isomorphism between MAGs and directed graphs, presented in Theorem~\ref{theo:iso}.

Given a MAG $H$, and by using the identity as the bijection $f$ used in Theorem~\ref{theo:iso}, we obtain the directed graph $G_n = (\mathbb{V}(H), E(G))$, where $E(G) \subseteq \mathbb{V}(H) \times \mathbb{V}(H)$. Note that $G_n$ is isomorphic to $H$ and corresponds to the natural isomorphism of $H$. In this case, we have that the function $h$ related to the isomorphism between $E(H)$ and $E(G)$~(see Corollary~\ref{coro:bijct}, Equation~\ref{func:h}) is written as
\begin{align}
h_i:E(H) & \to E(G) \\
(a_1,a_2,\dots,a_p,  b_1,b_2,\dots,b_p) & \mapsto ((a_1,a_2,\dots,a_p), (b_1,b_2,\dots,b_p)), \notag
\end{align}
where $E(G) \subseteq \mathbb{V}(H) \bigtimes \mathbb{V}(H)$. Here, $h_i$ represents the particular case of $h$ obtained by the natural isomorphism between $H$ and $G$.

Therefore, for any given MAG $H = (A, E)$, we can now define the function
\begin{align}
g:((A(H),E(H))) & \to (\mathbb{V}(H), E(G)) \\
H & \mapsto (I(\mathbb{V}(H)), h_i(E(H))), \notag
\end{align}
such that $g(H)$ is the composite vertices representation of the MAG $H$.
Since the graph $g(H)$ is the same graph obtained by the natural isomorphism shown in Section~\ref{subsec:magiso}, it follows that $g(H)$ is isomorphic to the MAG $H$.
Note that since a MAG is isomorphic to a traditional directed graph, it follows that a sub-MAG is a notion equivalent to the notion of sub-graphs in traditional directed graphs.

\subsubsection{Order preserving}
\label{subsubsec:ord}
From the MAG definition, aspects are not required to be ordered sets. As a consequence, the isomorphism defined in Theorem~\ref{theo:iso} does not necessarily preserves aspect ordering. If, however, a sublist of aspects on a MAG can support order, it is possible to obtain an isomorphism to a directed graph that preserves this order. 

Consider a MAG $H$ of order $p$, such that $A(H)$ (\emph{i.e.}, the aspect list of $H$) has a sublist $\zeta_o = (o_1,o_2,...,o_n)$ where the set $o_1 \times o_2 \times ... \times o_n$ admits order. Note that the sublist $\zeta_o$ characterizes a sub-determination of the MAG $H$. If we now consider the composite vertex set $\mathbb{V}_{\zeta_o}(H)$ of the sub-determined MAG $H$, it follows that this vertex set admits order, since $\mathbb{V}_{\zeta_o}(H) = o_1 \times o_2 \times ... \times o_n$. 

Since by construction each composite vertex of $H$ has exactly one element of each set in $\zeta_o$, it follows that we can partition the set $\mathbb{V}(H)$ by the elements of $\mathbb{V}_{\zeta_o}(H)$, so that $\mathbb{T}(H) = \mathbb{V}(H) / \mathbb{V}_{\zeta_o}(H)$ is the set of equivalence classes induced by $\zeta_o$ in  $\mathbb{V}(H)$. Therefore, as each equivalence class of $\mathbb{T}(H)$ has exactly one element of $\mathbb{V}_{\zeta_o}(H)$, it follows that $\mathbb{T}(H)$ can be ordered in the same way as $\mathbb{V}_{\zeta_o}(H)$.

This property can be useful for MAGs that, for instance, have one aspect that represents time instants. In this case, the MAG can be ordered in time by making each time instant correspond to an equivalence class.

\subsection{Degree}
\label{subsec:degree}
The concept of degree in a graph is associated with the concept of vertex. 
In a MAG, it is associated with an aspect, of which, as per our convention, a vertex is a special case. We therefore, define aspect and composite vertex degrees. 
Since the edges on a MAG are naturally directed, we adopt the same notation as in directed graphs of the indegree of a vertex~$u$ denoted as $deg^-(u)$ and the outdegree of a vertex~$u$ denoted as $deg^+(u)$. 

\subsubsection{Aspect degree}
\label{subsubsec:aspdeg}
We define the aspect degree as the number of edges incident to a given element of an aspect. Since the edges are directed, we distinguish between indegree and outdegree. The formal definition of aspect degree can therefore, be written as
\begin{align}
deg^+(a_i) & = | \{ e \in E(H) : \pi_i(e) = a_i \}|, \\
deg^-(a_i) & = | \{ e \in E(H) : \pi_{p+i}(e) = a_i \}|,
\end{align}
where $a_i \in A(H)[i]$ is an element of the $i$-th aspect of the MAG $H$, and $\pi_i$ is the canonical projection onto aspect $i$ defined in the beginning of Section~\ref{sec:mag}. Therefore, $deg^+(a_i)$ is the number of edges originated at element~$a_i$ and $deg^-(a_i)$ is the number of edges destined to element~$a_i$, noting that $a_i \in A(H)[i]$.

\subsubsection{Composite vertex degree}
\label{subsubsec:locdeg}
We also consider the degree based on composite vertices.
The composite vertex degree considers the degree of a vertex taking into account all the aspects present on the MAG. This follows directly from the definition of a composite vertex:
\begin{align}
deg^+(\mathbf{u}) & = | \{ e \in E(H) : \pi_o(e) = \mathbf{u} \}|, \\ 
deg^-(\mathbf{u}) & = | \{ e \in E(H) : \pi_d(e) = \mathbf{u} \}|.
\end{align}
That is, $deg^+(\mathbf{u})$ is the number of edges originated at the composite vertex $\mathbf{u}$, while $deg^-(\mathbf{u})$ is the number of edges destined to the composite vertex $\mathbf{u}$. 

\subsection{Predecessor and Successor}
\label{subsec:adj}

Given an edge $e \in E(H)$ for a MAG $H$, we say that the composite vertex $\mathbf{u} = \pi_o(e)$ is the predecessor of $\mathbf{v} = \pi_d(e)$, or that $\mathbf{v} = \pi_d(e)$ is the successor of $\mathbf{u} = \pi_o(e)$. Note that the same is valid for sub-determined vertices and edges. 

We say that two edges are adjacent if they have exactly one composite vertex in common, \emph{i.e.}, if there are two distinct edges $e_a,e_b \in E(H)$ and there is a composite vertex $\mathbf{u} \in \mathbb{V}(H)$, such that $\mathbf{u} \in \{\pi_o(e_a), \pi_d(e_a)\}$ and $\mathbf{u} \in \{\pi_o(e_b), \pi_d(e_b)\}$.

\begin{theorem}{Given a MAG $H$ and a directed graph $G$ isomorphic to $H$ and a pair of composite vertices $\mathbf{u}, \mathbf{v} \in \mathbb{V}(H)$, it follows that  $\mathbf{u}$ is the predecessor of $\mathbf{v}$ if and only if their corresponding vertices in~$G$ hold the same relation.}
\label{theo:adjlocnd}
\begin{proof}
Since $\mathbf{u}$ is the predecessor of $\mathbf{v}$ it follows that there is an edge $e \in E(H)$ such that $\mathbf{u} = \pi_o(e)$ and $\mathbf{v} = \pi_d(e)$. 
From Corollary~\ref{coro:bijct} we have that there is an edge $s = (f(\pi_o(e)), f(\pi_d(e))) \in E(G)$, where $f$ is a bijection from $\mathbb{V}(H)$ to $V(G)$, so that $f(\pi_o(e)) \in V(G)$ corresponds to $\mathbf{u}$ and $f(\pi_d(e)) \in V(G)$ corresponds to $\mathbf{v}$. Hence, as $f$ is a bijection, the theorem holds. 
\end{proof}
\end{theorem}

\begin{theorem}{Let $H$ be a MAG of order $p$, $\zeta$ a sub-determination, and $e_a, e_b \in  E(H)$, two distinct edges such that $S_\zeta(\pi_o(e_a)) \neq S_\zeta(\pi_d(e_a))$ and $S_\zeta(\pi_o(e_b)) \neq S_\zeta(\pi_d(e_b))$ and $E_\zeta(e_a) \neq E_\zeta(e_b)$ . If $e_a$ and $e_b$ are adjacent edges in $H$, then the sub-determined edges $E_\zeta(e_a)$ and $E_\zeta(e_b)$ are adjacent  on the sub-determined MAG $M_\zeta(H)$.}
\label{theo:agg}
\begin{proof}
Since $e_a$ and $e_b$ are adjacent edges in $H$, it follows that they share a common composite vertex, and are therefore, incident to three distinct composite vertices. Let $\mathbf{u} \in \mathbb{V}(H)$ be the shared composite vertex, and $\mathbf{v},\mathbf{w} \in \mathbb{V}(H)$ the other two composite vertices to which $e_a$ and $e_b$ are incident. Without loss of generality, we can assume that $\mathbf{u} = \pi_d(e_a) = \pi_o(e_b)$, $\mathbf{v} = \pi_o(e_a)$ and $\mathbf{w} = \pi_d(e_b)$. Since $E_\zeta(e_a) \neq E_\zeta(e_b)$, it follows that they are edges on $M_\zeta(H)$, \emph{i.e.} $E_\zeta(e_a), E_\zeta(e_b) \in \mathbb{E}\zeta(H)$. Further, $S_\zeta(\mathbf{u}) = S_\zeta(\pi_d(e_a)) = S_\zeta(\pi_o(e_b))$, $S_\zeta(\mathbf{v}) = S_\zeta(\pi_o(e_a))$ and $S_\zeta(\mathbf{w}) = S_\zeta(\pi_d(e_b))$. Therefore, $S_\zeta(\mathbf{u})$, $S_\zeta(\mathbf{v})$ and $S_\zeta(\mathbf{w})$ are three distinct composite vertices on $M_\zeta(H)$, so that $E_\zeta(e_a)$ and $E_\zeta(e_b)$ are adjacent, since $S_\zeta(\mathbf{u})$ is a composite vertex shared by them.
\end{proof}
\end{theorem}

\subsection{Walks, trails, paths, and cycles}
\label{subsec:wtpc}

We define a walk on a MAG $H$ of order $p$ as an alternating sequence $W = [\mathbf{u}_{1},e_1,\mathbf{u}_{2},$ $e_2,\mathbf{u}_{3},\dots,\mathbf{u}_{{k-1}}$, $e_{k-1},\mathbf{u}_{k}]$ of composite vertices $\mathbf{u}_{n} \in \mathbb{V}(H)$ and edges $e_m \in E(H)$, such that $\mathbf{u}_{n} = \pi_o(e_n)$ and $\mathbf{u}_{{n+1}} =   \pi_d(e_n)$ for $1 \leq n < k$.
Note that from this definition we have that for all pairs of consecutive composite vertices $\mathbf{u}_{m}$ and $\mathbf{u}_{{m+1}}$, $1 \leq m \leq k$, are adjacent and also that for all pairs of consecutive edges $e_j$ and $e_{j+1}$,$1 \leq j < k$, are adjacent as well.

A walk is closed if $\mathbf{u}_{1} = \mathbf{u}_{k}$ and open otherwise. The set of composite vertices in the walk $W$ is denoted as ${V}(W)$ and the set of edges in the walk $W$ is denoted as $E(W)$. Since the edges in $W$ contain elements of every aspect in $A(H)$, it follows that a walk has the same $p$ aspects of the MAG $H$ where the walk is defined, \emph{i.e.} $|A(W)| = |A(H)|$. However, for each aspect $A(W)[n] \in A(W)$, we have that $A(W)[n] \subseteq A(H)[n]$, since not necessarily each element of a given aspect will be reached by the walk.
Therefore, if $W$ is a walk on a MAG~$H$, then ${V}(W) \subseteq \mathbb{V}(H)$, $E(W) \subseteq E(H)$,  $|A(W)| = |A(H)|$, and $A(W)[n] \subseteq A(H)[n]$ for $1 \leq n \leq p$, where $p = |A(W)| = |A(H)|$ is the order of both $W$ and $H$.

Note that each edge $e_n$ in a walk $W$ can be determined from the composite vertices $\mathbf{u}_{n}$ and $\mathbf{u}_{{n+1}}$ by noting that $\mathbf{u}_{n}$ is the predecessor of $\mathbf{u}_{{n+1}}$.
Therefore, $W$ can be fully described by the sequence of its composite vertices, $W_V = [\mathbf{u}_{1}, \mathbf{u}_{2}, ..., \mathbf{u}_{k}]$. We may refer to a walk using this notation in cases where the precise determination of the edges is not needed. The sequence of composite vertices $W_V$ is not necessarily equal to the set ${V}(W)$ of composite vertices in the walk, since in $W_V$ there may be repeated composite vertices.

Further, each edge $e_j$ in a walk $W$ also fully determines the composite vertices $\mathbf{u}_{j}$ and $\mathbf{u}_{{j+1}}$, since $\mathbf{u}_{j} = \pi_o(e_j)$ and $\mathbf{u}_{{j+1}} = \pi_d(e_j)$. Hence, $W$ can also be determined by its sequence of edges $W_E = [e_1, e_2, \dots,$ $e_{k-1}]$. We may use this notation when the precise identification of the composite vertices is not needed. The sequence of edges $W_E$ is not necessarily equal to the set of edges $E(W)$, since there may be repeated edges in $W_E$. The length of a walk is determined by the number of edges the walk contains, \emph{i.e.} $Len(W) = |W_E|$.

As a short notation, in cases where there is no ambiguity, or the identity of the composite vertices and edges in the walk is irrelevant, we may also identify a walk $W$ only by its starting and ending composite vertices as $W = \mathbf{u}_{1} \to \mathbf{u}_{k}$.

We define a trail in a MAG $H$ as a walk on $H$ where all edges are distinct. Since all edges are distinct, we can identify a trail $W = [\mathbf{u}_{1},e_1,\mathbf{u}_{2},e_2,\mathbf{u}_{3},...,\mathbf{u}_{{k-1}}$, $e_{k-1},\mathbf{u}_{k}]$ with the MAG $H_W$ = (A(W), E(W)), where $E(W) \subseteq E(H)$, $|A(W)| = |A(H)|$, and $A(W)[n] \subseteq A(H)[n]$ for $1 \leq n \leq p$, where $p = |A(W)|$ is the order of $W$. Therefore, the trail W is a sub-MAG of $H$.
A trail is closed when the first and last composite vertices are the same, \emph{i.e.} $\mathbf{u}_{1} = \mathbf{u}_{k}$, and open otherwise. A closed trail is also called a tour or a circuit. 

We define a path on a MAG $H$ as a walk on $H$ where all composite vertices are distinct.
We can associate a path $P = [\mathbf{u}_{1},e_1,\mathbf{u}_{2},e_2,\mathbf{u}_{3},...,\mathbf{u}_{{k-1}}$, $e_{k-1},\mathbf{u}_{k}]$ with the MAG $H_P$ = (A(P), E(P)), where $E(P) \subseteq E(H)$,  $|A(P)| = |A(H)|$, and $A(P)[n] \subseteq A(H)[n]$ for $1 \leq n \leq p$, where $p = |A(H)|$ is the order of $H$. Since all composite vertices in $P$ are distinct, it follows that all edges in $P$ are also distinct, because each edge in $P$ is determined by the two composite vertices adjacent to it. Therefore, we have that the path~$P$ is a sub-MAG of $H$. 

\begin{proposition}{Walks, Trails, Paths, and Cycles}
\label{prop:walks}\\
Theorem~\ref{theo:iso}  assures that every MAG $H$ has a unique directed graph $G$ isomorphic to it. From Corollary~\ref{coro:bijct}, it follows that every edge in a MAG $H$ has a unique edge associated with it in the $G$. Finally, from Theorem~\ref{theo:adjlocnd}, it follows that the isomorphism between MAGs and directed graphs preserves the predecessor and successor relation of vertices.

As a direct consequence of these theorems, the following propositions hold.

\begin{enumerate}
\item An alternating sequence $W$ of composite vertices and edges in a MAG~$H$ is a walk on $H$ if and only if there is a corresponding walk $G_W$ in the composite vertices representation of~$H$.
\label{wk}

\item The length of a walk $W$ on a MAG $H$ is the same as the length of the corresponding walk $G_W$ on the directed graph $g(H)$.
\label{lwk}

\item A walk $H_W$ on a MAG $H$ is a trail on $H$ if and only if there is a corresponding trail $G_W$ on $g(H)$.
\label{wt}

\item The length of a trail $H_W$ on a MAG $H$ is the same as the length of the corresponding trail $G_W$ on $g(H)$, \emph{i.e.} $Len(H_W) = Len(G_W)$.
\label{lwt}

\item A walk $P$ on $H$ is a path on $H$ if and only if there is a corresponding path $G_P$ on $g(H)$.
\label{wp}

\item The length of a path $P$ on a MAG $H$ is the same as the length of the corresponding path $G_P$ on $g(H)$, \emph{i.e.} $Len(P) = Len(G_P)$.
\label{lwp}

\item A path $P$ on a MAG $H$ is a cycle if and only if the corresponding path $G_P$ in $g(H)$ is a cycle.
\label{pc}
\end{enumerate}
\end{proposition}

\begin{theorem}{Given a MAG $H$ and a sub-determination $\zeta$, the projection of a walk on $H$ onto $M_\zeta(H)$ corresponds to a walk on $M_\zeta(H)$.}
\label{theo:aggwalk}
\begin{proof}
Let $H$ be a MAG of order $p$, $M_\zeta(H)$ a sub-determined MAG of $H$ and $W = [\mathbf{u}_{1},e_1,$ $\mathbf{u}_{2},e_2,\mathbf{u}_{3},...,\mathbf{u}_{{k-1}}$, $e_{k-1},\mathbf{u}_{k}]$ a walk on MAG $H$. Consider $W_\zeta = [S_\zeta(\mathbf{u}_{1}),E_\zeta(e_1),S_\zeta(\mathbf{u}_{2}),E_\zeta(e_2),S_\zeta(\mathbf{u}_{3}),...,S_\zeta(\mathbf{u}_{{k-1}}), E_\zeta(e_{k-1}),S_\zeta(\mathbf{u}_{k})]$.
If all the edges $e_n \in E(W)$ are such that $S_\zeta(\pi_o(e_n)) \neq S_\zeta(\pi_d(e_n))$ and $E_\zeta(e_n) \neq E_\zeta(e_{n+1})$, then every consecutive pair of composite vertices $S_\zeta(\mathbf{u}_{n})$ and $S_\zeta(\mathbf{u}_{n+1})$ are distinct and this theorem holds as a direct consequence of Theorem~\ref{theo:agg}. If for a given edge $e_n \in E(H)$ we have that $S_\zeta(\pi_o(e_n)) = S_\zeta(\pi_d(e_n))$, this means that $E_\zeta(e_n)$ is a self-loop. In this case, $E_\zeta(e_n)$ and $S_\zeta(\mathbf{u}_{n+1})$ are dropped from $W_\zeta$, eliminating the self-loop. If for a given pair of consecutive edges $e_n, e_{n+1}$ we have that $E_\zeta(e_n) = E_\zeta(e_{n+1})$, then both $E_\zeta(e_n)$ and $E_\zeta(e_{n+1})$ are self-loops, so that $E_\zeta(e_n)$, $S_\zeta(\mathbf{u}_n$, $E_\zeta(e_{n+1})$ and $S_\zeta(\mathbf{u}_{n+1})$ are dropped from $W_\zeta$. Once all self-loops are removed from $W_\zeta$, the remaining alternating sequence of composite vertices and edges is a walk on $M_\zeta(H)$ and therefore, this theorem holds. Note, however, that $W_\zeta$ may be reduced to a single composite vertex and no edge.
\end{proof}
\end{theorem}
Since Theorem~\ref{theo:aggwalk} holds for any given sub-determination, it also holds for the aggregated graph of $H$, since it is a special case of sub-determination.

From Theorem~\ref{theo:aggwalk}, we have that the projection of a trail $W$ onto a sub-determined MAG $M_\zeta(H)$ is a walk, \emph{i.e.} the projection of a trail does not necessarily lead to a trail on $M_\zeta(H)$. Consider the trail $W = [\mathbf{u}_1,(\mathbf{u}_1, \mathbf{u}_2), \mathbf{u}_2, (\mathbf{u}_2, \mathbf{u}_3),$ $\mathbf{u}_3, (\mathbf{u}_3, \mathbf{u}_4),$ $ \mathbf{u}_4, (\mathbf{u}_4, \mathbf{u}_5), \mathbf{u}_5]$, where all five composite nodes are distinct, but $S_\zeta(\mathbf{u}_1)$  $= S_\zeta(\mathbf{u}_4)$ and  $S_\zeta(\mathbf{u}_2) = S_\zeta(\mathbf{u}_5)$. Note that $E_\zeta((\mathbf{u}_1, \mathbf{u}_2)) = E_\zeta((\mathbf{u}_4, \mathbf{u}_5))$ and, therefore, $W_\zeta$ is not a trail on $M_\zeta$.

From Theorem~\ref{theo:aggwalk}, we have that the projection of a path $W$ onto a sub-determined MAG $M_\zeta(H)$ is a walk. We intend to show that such projection of a path does not necessarily lead to a trail on $M_\zeta(H)$. Consider the path $W = [\mathbf{u}_1,(\mathbf{u}_1, \mathbf{u}_2), \mathbf{u}_2, (\mathbf{u}_2, \mathbf{u}_3),$ $\mathbf{u}_3, (\mathbf{u}_3, \mathbf{u}_4), \mathbf{u}_4, (\mathbf{u}_4, \mathbf{u}_5), \mathbf{u}_5]$, where all five composite nodes are distinct, but $S_\zeta(\mathbf{u}_1) = S_\zeta(\mathbf{u}_5)$. Since all composite nodes of $W$ are distinct, $W$ is a path on the MAG $H$. Let's consider the projection onto $M_\zeta$, $W_\zeta = [S_\zeta(\mathbf{u}_1), E_\zeta((\mathbf{u}_1, \mathbf{u}_2)), S_\zeta(\mathbf{u}_2),$ $E_\zeta((\mathbf{u}_2, \mathbf{u}_3)), S_\zeta(\mathbf{u}_3), E_\zeta((\mathbf{u}_3, \mathbf{u}_4)), S_\zeta(\mathbf{u}_4), E_\zeta((\mathbf{u}_4, \mathbf{u}_5)), S_\zeta(\mathbf{u}_5)]$. Since $S_\zeta(\mathbf{u}_1) = S_\zeta(\mathbf{u}_5)$, there is a repeated composite vertex in $W_\zeta$, and therefore, $W_\zeta$ is not a path on $M_\zeta$.

\subsection{Shortest paths on a MAG}
\label{subsec:shortph}
Before defining a shortest path in a MAG, we first present the concept of shortest walk.
Given two distinct composite vertices $\mathbf{u}_{a}, \mathbf{u}_{b} \in \mathbb{V}(H)$ in a MAG $H$, such that there is at least one walk from $\mathbf{u}_{a}$ to $\mathbf{u}_{b}$,
we define the shortest walk between $\mathbf{u}_{a}$ and $\mathbf{u}_{b}$ as the walk $W_s$ from $\mathbf{u}_{a}$ to $\mathbf{u}_{b}$ such that no other walk from $\mathbf{u}_{a}$ to $\mathbf{u}_{b}$ is shorter than~$W_s$. Note that a shortest walk between a given pair of distinct composite vertices is not necessarily unique.

\begin{theorem}{A shortest walk $W_s$ between a given pair of distinct composite vertices on a MAG is necessarily a path.}
\label{theo:shortwalk}
\begin{proof} (by contradiction) \\
Let $H$ be a MAG of order $p$, $\mathbf{u}_{1},\mathbf{u}_{q} \in \mathbb{V}(H)$ two distinct composite vertices in $H$, and $P = [\mathbf{u}_{1}, \mathbf{u}_{2}, \dots, \mathbf{u}_{q}]$ a shortest walk from $\mathbf{u}_{1}$ to $\mathbf{u}_{q}$ in $H$. We now assume that $P$ is not a path (\emph{i.e.} $P$ has at least one repeated composite vertex). Then, there is (at least) a pair of composite vertices $\mathbf{u}_{j},\mathbf{u}_{k} \in P$, such that $\mathbf{u}_{j} = \mathbf{u}_{k}$, $j \neq k$, and $j < k$. In this case, $P_s = [\mathbf{u}_{1}, \mathbf{u}_{2}, \dots, \mathbf{u}_{j}, \mathbf{u}_{{k+1}}, \dots, \mathbf{u}_{q}]$ is a walk from $\mathbf{u}_{1}$ to $\mathbf{u}_{q}$ which is shorter than $P$. This is a contradiction, since $P$ is a shortest walk from $\mathbf{u}_{1}$ to $\mathbf{u}_{q}$ in $H$.

Note that the same argument holds for particular cases, such as when $j = 1$ or $k = q$. We can therefore conclude that $P$ is a path, and the theorem holds.
\end{proof}
\end{theorem}

\begin{theorem}{A path $P$ between two composite vertices $\mathbf{u}$ and $\mathbf{v}$ on a MAG $H$ is a shortest path between these vertices if and only if the corresponding path $G_P$ in $g(H)$ is a shortest path between the vertices $\mathbf{u}$ and $\mathbf{v}$ in $g(H)$.}
\label{theo:short}
\begin{proof}
\begin{itemize}
\item $\Longrightarrow$ (by contradiction) \\
Let $\mathbf{u}, \mathbf{v} \in \mathbb{V}(H)$ be two composite vertices on a MAG $H$ and $P$ be a shortest path from $\mathbf{u}$ to  $\mathbf{v}$. Further, let $G_P$ be the corresponding path from $\mathbf{u}$ to  $\mathbf{v}$ on the graph $g(H)$. From Proposition~\ref{prop:walks} (\ref{lwp}), we have that $Len(P) = Len(G_P)$. \\
Let's suppose that $G_P$ is not a shortest path from $\mathbf{u}$ to  $\mathbf{v}$ on $g(H)$. This means that there is a path $G_{P_S}$ from $\mathbf{u}$ to  $\mathbf{v}$ on $g(H)$, such that $Len(G_{P_S}) < Len(G_P)$. Then, by Proposition~\ref{prop:walks} (\ref{wp})~and~(\ref{lwk}), there must be a corresponding path $P_S$ from $\mathbf{u}$ to  $\mathbf{v}$ on $H$, such that $Len(P_S) < Len(P)$. This is a contradiction, since $P$ is a shortest path from $\mathbf{u}$ to  $\mathbf{v}$ on $H$. Therefore, $G_P$ is a shortest path from $\mathbf{u}$ to  $\mathbf{v}$ on $g(H)$.
\item $\Longleftarrow$ (by contradiction) \\
Let $\mathbf{u}, \mathbf{v} \in V(g(H))$ be two vertices on $g(H)$ and $G_P$ be a shortest path from $\mathbf{u}$ to  $\mathbf{v}$. Further, let $P$ be the corresponding path from $\mathbf{u}$ to  $\mathbf{v}$ on the MAG $H$. From Proposition~\ref{prop:walks} (\ref{lwp}), we have that $Len(G_P) = Len(P)$. \\
Let's suppose that $P$ is not a shortest path from $\mathbf{u}$ to  $\mathbf{v}$ on MAG $H$. Then, there must be a path $P_S$ from $\mathbf{u}$ to  $\mathbf{v}$ in $H$, such that $Len(P_S) < Len(P)$. Thus, from Proposition~\ref{prop:walks} (\ref{wp})~and~(\ref{lwk}), there must be a corresponding path $G_{P_S}$ from $\mathbf{u}$ to  $\mathbf{v}$ on $g(H)$, such that $Len(G_{P_S}) < Len(G_P)$. This is a contradiction, since $G_P$ is a shortest path from $\mathbf{u}$ to  $\mathbf{v}$ on $g(H)$. Therefore, $P$ is a shortest path from $\mathbf{u}$ to  $\mathbf{v}$ on the MAG $H$.
\end{itemize}
\end{proof}
\end{theorem}

\section{MAG applicability}
\label{sec:app}

The MAG concept is a graph generalization able to represent multilayer and time-varying networks, as well as time-varying multilayer networks.
Aspects in a MAG represent key independent features of the complex networked system to be represented, such as time instants or layers, generalizing
the notion of vertex. In this section, we thus present an overview of the applicability of the proposed MAG abstraction. 
Table~\ref{tab:use} summarizes the possible MAG applicability as a function of the MAG order $|A|$, \emph{i.e.} the number of aspects present in each case.
Order $1$ MAGs are traditional directed graphs, order $2$ MAGs can represent time-varying graphs (TVGs) or multilayer graphs, order $3$ MAGs can represent objects like time-varying multilayer graphs, and so on. 

\begin{table}[!h]
\centering
\caption{MAG applicability as a function of the MAG order $|A|$.}
\label{tab:use}
\begin{tabular}{c c c c c c}
\hline \hline
$|A|$	& Composite Vertex    & \hspace*{0.2cm}	& Edge 				& \hspace*{0.2cm}	&	Examples		 		\\	
\hline \hline
1			& single object 		& \hspace*{0.2cm}	&  ordered pair			& \hspace*{0.2cm}	& 	directed graph			\\
2			& ordered pair		& \hspace*{0.2cm}	&  ordered quadruple	& \hspace*{0.2cm}	& TVG or multilayer		\\
3			& ordered triple		& \hspace*{0.2cm}	&  ordered sextuple		& \hspace*{0.2cm}	& Time-varying multilayer		\\
4			& ordered quadruple	& \hspace*{0.2cm}	&  ordered octuple		& \hspace*{0.2cm}	& \dots						\\
\hline \hline
\end{tabular}
\end{table}

The remainder of this section discusses some general examples of the applicability of order $2$ and $3$ MAGs as well as of the basic MAG representation and algorithms.

\subsection{Multilayer graphs}
\label{subsec:multi}
Multilayer graphs are used to represent networked systems where distinct complex networks interact with each other and can be represented
as a layered system~\cite{Kurant2006a}.
Examples of such systems are, for instance, power supply networks, which have distinct power and control networks~\cite{Buldyrev2010}, or an arrangement of multiple online social networks~(such as facebook, linkedin, and twitter) where users with accounts on multiple networks act as an interconnection between them.

Note that this kind of multilayer networked systems can be represented by using order $2$ MAGs, with one aspect for the vertices and another aspect for the layers, which are the key independent features of this kind of systems. For instance, in the case of power supply networks, these networks could be represented by a MAG with aspects ``device" and ``layer", while the multiple online social networks can be represented by an order $2$ MAG with aspects ``user" and ``online social network".

\subsection{Time-varying graphs (TVGs)}
\label{subsec:tvg}
Recently, there is an increasing interest in studying time-varying networks, which can be seen as networks whose structure (vertices and edges) may vary in time. As a consequence of this interest, a number of distinct representations for this kind of network has been proposed~\cite{Casteigts2012,Holme2012,Ferreira2004,Kostakos2009,Tang2010a}. Some of these representations are equivalent. However, some of them are not, \emph{i.e.}, it is not possible to directly translate a network from a model to another one. In~\cite{Wehmuth2015a}, we show that, by using a TVG constructed with a MAG model~(\emph{i.e.}, a order $2$ MAG), we derive an unifying model that can represent several previous (classes of) models for time-varying networks found in the recent literature, which in general are unable to represent each other. In particular, the previous classes of models for TVG representation that are shown in~\cite{Wehmuth2015a} to be unified by a MAG-based TVG representation include TVG models based on: (i)~snapshots, such as those used by~\cite{Holme2012,Ferreira2004,Tang2010a}; (ii)~continuous time intervals~\cite{Casteigts2012}; (iii)~spatial and temporal edges~\cite{Kostakos2009}; and (iv)~temporal and mixed edges~\cite{Kim2012}. The fundamental characteristic that allows the MAG-based TVG representation proposed in~\cite{Wehmuth2015a} to unify these previous (classes of) models is that it has a set of types of edges that is able to represent all the previous models. More specifically, an edge $e= (u, t_a, v, t_b)$, where $u$ and $v$ are vertices while $t_a$ and $t_b$ are time instants, in the MAG-based TVG model may be classified into four classes depending on its temporal characteristic:
\begin{enumerate}
\item \emph{Spatial edges} connect two nodes at the same time instant, $e$ is in the form of  $e =(u, t_a, v, t_a)$, where $u \neq v$;
\item \emph{Temporal edges} connect the same node at two distinct time instants, $e$ is in the form of $e=(u, t_a, u, t_b)$, where $t_a \neq t_b$;
\item \emph{Mixed edges} connect distinct nodes at distinct time instants, $e$ is in the form of $e=(u, t_a, v, t_b)$, where $u \neq v$ and $t_a \neq t_b$;
\item \emph{Spatial-temporal self-loop edges} connect the same node at the same time instant, $e$ is in the form of $e=(u, t_a, u, t_a)$.
\end{enumerate}

Each of the previous TVG models actually uses a subset of these types of edges. In particular, for the TVG models based on continuous time intervals, a discretization process is applied in order to have a discrete construction that uses a subset of these types of edges. In a similar way, the use of different subsets of these types of edges is why the previous models are unable to represent each other.  In contrast, since the MAG-based TVG representation allows for the use of all these types of edges, it is able to be a unifying model for previous (classes of) models for TVGs, which are unable to represent each other. We refer the interested reader to~\cite{Wehmuth2015a} for further details and examples on this unifying MAG-based representation for TVG models.

Building upon this MAG-based unifying model for TVGs, Costa~et~al.~\cite{EduardoACS2015} introduce and investigate the notion of time centrality to evaluate the relative importance of time instants in TVGs.  It is shown that diffusion starting at the best ranked time instants (\emph{i.e.} the most central ones), according to the considered metrics, can perform a faster and more efficient diffusion process in TVGs.

Due to the time-varying characteristic of TVGs, in some cases it can be of interest to determine distances in terms of time. This kind of temporal distance can be extracted from a MAG representation of a time-varying network. In particular, in a MAG where the time instants are ordered, and all edges with a time component follow the increasing order of time, the temporal distance of a path can be obtained from the sub-determination of the path upon the time aspect (\emph{i.e.} representing the path only on the time aspect). 
Note that a temporal or mixed edge $e = (u, t_a, v, t_b)$ can be progressive~($t_a < t_b$) or regressive~($t_a > t_b$) in time. Progressive temporal edges represent the intuitive evolution in time of the TVG. In contrast, regressive temporal edges can intrinsically model cyclic (i.e., periodic) behavior in dynamic networks~\cite{Wehmuth2015a}. 


\subsection{MAG representation and algorithms}
\label{subsec:repalg}

In this paper, we introduce the MAG abstraction, briefly discussing its applicability in this section. For practical use, however, it is necessary to have ways to properly and efficiently represent MAGs. Furthermore, it is also necessary to have a set of basic algorithms capable of manipulating MAGs that can be used to build more advanced algorithms for the particular analysis of the application of interest. 

In this context, we provide in a companion paper~\cite{Wehmuth2015} discussion in further details such basic MAG representations and algorithms. 
In particular,
we present basic manipulation algorithms that can perform operations like build MAGs, extract sub-MAGs, and make MAG sub-determinations. As indicated at the end of Section~\ref{subsubsec:crep}, sub-MAGs are the equivalent of sub-graphs for MAGs. In contrast, a MAG sub-determination, defined in Section~\ref{subsubsec:subdetmag}, is similar to the concept of aggregation, which is well-known in the theory of  multilayer graphs and time-varying graphs.
Aggregations consist in projecting all edges upon a single layer of the graph and thereby transforming it into a traditional graph. In time-varying graphs, for instance, this corresponds to eliminate the notion of time, whereas in multilayer graphs an aggregation corresponds to eliminate the layers. 
However, in MAGs with order greater than 2, there are many complex ways to do aggregations. In fact, a sub-determination is a generalization of aggregation.
Further, we recall that Section~\ref{subsubsec:subdetvt} shows that an order $p$ MAG admits $2^p - 2$ distinct sub-determinations, coinciding with the fact that traditional graphs (order 1 MAGs) cannot be sub-determined. In short, we remark that a Sub-MAG reduces the number of composite vertices of the original MAG, whereas a sub-determination of the same MAG reduces its number of aspects.

Further, in~\cite{Wehmuth2015}, we also present more elaborated graph algorithms for MAGs, such as breadth-first search~(BFS) and depth-first search~(DFS), that can be used as building blocks for other uses, thus allowing the extension of the MAG applicability. 
Concerning algorithm complexity, given the isomorphism between a MAG and a traditional directed graph, we expect MAG algorithms to have the same complexity as the equivalent algorithms for traditional directed graphs. For instance, 
if the MAG algorithm is derived from a given traditional algorithm which is polynomial for traditional graphs, the resulting algorithm for MAGs will be polynomial in the size of the composite vertices representation of the MAG.

As discussed in~\cite{Wehmuth2015}, MAG algorithms are presented in two forms: (i)~the natural form, where the result is expressed in terms of full composite vertices; and (ii)~the sub-determinated form, where the result is given as sub-determinated vertices. 
Sub-determined algorithms are algorithms that receive a given MAG, its companion tuple, and a given sub-determination. The output of sub-determined algorithms is expressed in terms of sub-determined vertices, according to the sub-determination provided to the algorithm. At a first glance, this seems similar to sub-determine the MAG and run the natural form of the desired algorithm on the resulting sub-determined MAG. 
However, in this case, results may be affected by spurious paths potentially created by the sub-determination process, in a similar way to what may happen in a traditional aggregation process in time-varying or multilayer graphs. 
Since the sub-determined algorithm uses the original MAG, the results can be computed disregarding potentially spurious paths generated by a sub-determination, therefore always getting the correct results~\cite{Wehmuth2015}.



As a further contribution, we also make available Python implementations of all the algorithms presented in~\cite{Wehmuth2015} at the following URL:~\url{http://github.com/wehmuthklaus/MAG_Algorithms}.

\section{Conclusion}
\label{sec:conc}
In this paper, we have formalized the MultiAspect Graph~(MAG) concept and have proved that a MAG is isomorphic to a traditional directed graph.
This leads to an important theoretical framework because this allows the use of the isomorphic directed graph as a tool to analyze both the properties of a MAG and the behavior of dynamic processes over a MAG.
Further, we have also demonstrated that other key MAG properties, such as adjacency, walks, trails, and paths, are also closely related to their counterparts
in a traditional directed graph. Such demonstrated properties allow MAG-based models of complex networked systems to be analyzed with the help of directed graph arguments.
 
The MAG concept thus enables the modelling of networked objects with characteristics similar to traditional graphs, but that simultaneously also present some dependency on other \emph{aspects}, such as layers and/or time. Since the MAG structure admits an arbitrary (finite) number of aspects, it hence introduces a powerful modelling abstraction for higher-order networked complex systems. As future work, we intend to further investigate the MAG applicability, such as studying centrality analysis on higher-order networks using MAGs and
applying the MAG concept on the modeling of higher-order networked systems from different domains.  

\section*{Acknowledgment}
This work was partially funded by the Brazilian funding agencies CAPES (STIC-AmSud Program), CNPq (grant number 308.729/2015-3), FINEP (grant number 0615/11), and FAPERJ (grant number E-26/103.207/2011) as well as the Brazilian Ministry of Science, Technology, Innovations, and Communications (MCTIC).






\begin{thebibliography}{10}
\expandafter\ifx\csname url\endcsname\relax
  \def\url#1{\texttt{#1}}\fi
\expandafter\ifx\csname urlprefix\endcsname\relax\def\urlprefix{URL }\fi
\expandafter\ifx\csname href\endcsname\relax
  \def\href#1#2{#2} \def\path#1{#1}\fi

\bibitem{Distel2010}
R.~Distel, Graph Theory, 4th Edition, Springer, 2010.

\bibitem{Erdos1964}
P.~Erd\"{o}s, \href{http://link.springer.com/article/10.1007/BF02759942}{{On
  extremal problems of graphs and generalized graphs}}, Israel Journal of
  Mathematics 2~(3) (1964) 183--190.

\bibitem{Bollobas1965}
B.~Bollob\'{a}s, \href{http://www.akademiai.com/index/V84963547704K070.pdf}{{On
  generalized graphs}}, Acta Mathematica Hungarica 16~(3-4) (1965) 447--452.

\bibitem{Chvatal1971}
V.~Chvatal,
  \href{http://www.jstor.org/stable/2036470?origin=crossref}{{Hypergraphs and
  Ramseyian theorems}}, Proceedings of the American Mathematical Society 27~(3)
  (1971) 434.

\bibitem{Kurant2006a}
M.~Kurant, P.~Thiran,
  \href{http://link.aps.org/doi/10.1103/PhysRevLett.96.138701}{{Layered complex
  networks}}, Physical Review Letters 96~(13) (2006) 138701.

  \bibitem{DeDomenico2013}
M.~{De Domenico}, A.~Sol\'{e}-Ribalta, E.~Cozzo, M.~Kivel\"{a}, Y.~Moreno,
  M.~Porter, S.~G\'{o}mez, A.~Arenas,
  \href{http://link.aps.org/doi/10.1103/PhysRevX.3.041022}{{Mathematical
  formulation of multilayer networks}}, Physical Review X 3~(4) (2013) 041022.

\bibitem{Gomez2013}
S.~G\'{o}mez, A.~D\'{\i}az-Guilera, J.~G\'{o}mez-Garde\~{n}es, C.~J.
  P\'{e}rez-Vicente, Y.~Moreno, A.~Arenas,
  \href{http://link.aps.org/doi/10.1103/PhysRevLett.110.028701}{{Diffusion
  dynamics on multiplex networks}}, Physical Review Letters 110~(2) (2013)
  028701.
  
\bibitem{Kivela14072014}
M.~Kivel\"{a}, A.~Arenas, M.~Barthelemy, J.~P. Gleeson, Y.~Moreno, M.~A.
  Porter, Multilayer networks, Journal of Complex Networks (2014) 1--69.
  
 \bibitem{Casteigts2012}
A.~Casteigts, P.~Flocchini, W.~Quattrociocchi, N.~Santoro,
  \href{http://www.tandfonline.com/doi/abs/10.1080/17445760.2012.668546}{{Time-varying
  graphs and dynamic networks}}, International Journal of Parallel, Emergent
  and Distributed Systems 27~(5) (2012) 387--408.

\bibitem{Holme2012}
P.~Holme, J.~Saram\"{a}ki,
  \href{http://linkinghub.elsevier.com/retrieve/pii/S0370157312000841}{{Temporal
  networks}}, Physics Reports 519~(3) (2012) 97--125.
  
  \bibitem{flocchini2013}
P.~Flocchini, B.~Mans, N.~Santoro,
   {{On the exploration of time-varying networks}}, Theoretical Computer Science 469 (2013) 53--68.

\bibitem{holme2013}
P.~Holme, J.~Saram\"{a}ki (Eds.), Temporal Networks, Springer, 2013.

\bibitem{Wehmuth2015a}
K.~Wehmuth, A.~Ziviani, E.~Fleury, {{A
  unifying model for representing time-varying graphs}}, Proceedings of the 
  IEEE International Conference on Data Science and Advanced Analytics (DSAA), (2015).
  
\bibitem{Kim2012-mag}
M.~Kim, J.~Leskovec,
{{Multiplicative Attribute Graph Model of Real-World Networks}},
Internet Mathematics 8~(2) (2012).

\bibitem{Mucha2010b}
P.~J. Mucha, T.~Richardson, K.~Macon, M.~A. Porter, J.-P. Onnela,
  \href{http://www.ncbi.nlm.nih.gov/pubmed/20466926}{{Community structure in
  time-dependent, multiscale, and multiplex networks.}}, Science 328~(5980)
  (2010) 876--8.


\bibitem{Buldyrev2010}
S.~Buldyrev, R.~Parshani, G.~Paul et al.
  {{Catastrophic cascade of failures in interdependent networks}}, Nature 464~(7291)
  (2010) 1025--8.
  
\bibitem{Ferreira2004}
A. Ferreira,
{{Building a reference combinatorial model for MANETs}},
IEEE Network 18~(5) (2004) 24--29.

\bibitem{Kostakos2009}
V.~Kostakos.
  {{Temporal graphs}}, Physica A: Statistical Mechanics and its Applications 388~(6)
  (2009) 1007--1023.

\bibitem{Tang2010a}
J.~Tang, S.~Scellato, M.~Musolesi, C.~Mascolo, V.~Latora,
{{Small-world behavior in time-varying graphs}}, 
Physical Review E 81~(5) (2010) 81--84.

\bibitem{Kim2012}
H.~Kim, R.~Anderson,
{{Temporal node centrality in complex networks}},
Physical Review E 85~(2) (2012).	

\bibitem{EduardoACS2015}
E.~C.~Costa, A.~B.~Vieira, K.~Wehmuth, A.~Ziviani, A.~P.~C.~Silva, {Time centrality in 
dynamic complex networks}, Advances in Complex Systems (ACS), 18~(07n08) (2015) 1550023.

\bibitem{Wehmuth2015}
K.~Wehmuth,  E.~Fleury, A.~Ziviani, {{MultiAspect Graphs: Algebraic representation and algorithms}}, 
arXiv:1504.07893, (2015) 1--61.

%
%
%
%
%
%
\end{thebibliography}
\end{document}